# Adjustable spin torque in magnetic tunnel junctions with two fixed layers

G. D. Fuchs, I. N. Krivorotov, P. M. Braganca, N. C. Emley, A. G. F. Garcia, D. C. Ralph, and R. A. Buhrman
*Cornell University, Ithaca, New York 14853-2501*

We have fabricated nanoscale magnetic tunnel junctions (MTJs) with an additional fixed magnetic layer added above the magnetic free layer of a standard MTJ structure. This acts as a second source of spin-polarized electrons that, depending on the relative alignment of the two fixed layers, either augments or diminishes the net spin-torque exerted on the free layer. The compound structure allows a quantitative comparison of spin-torque from tunneling electrons and from electrons passing through metallic spacer layers, as well as analysis of Joule self-heating effects. This has significance for current-switched magnetic random access memory (MRAM), where spin torque is exploited, and for magnetic sensing, where spin torque is detrimental.

Recent demonstrations[1,2] of magnetic random access memory (MRAM) have shown that magnetic memory may compete with silicon based memory in some applications. In order to scale cell size and decrease write currents, future generations of MRAM may use spin transfer as the write mechanism[3,4]. A key to spin-transfer switched MRAM is the use of magnetic tunnel junctions[5,6] (MTJs) to obtain an impedance match to CMOS technology. Initial demonstrations of spin transfer in MTJs indicate critical current densities similar to those observed in current perpendicular to the plane (CPP) spin valves[3], but with higher bias voltages (~0.5 V). This poses questions regarding self-heating effects and the possible degradation and breakdown of the ultra-thin insulating barrier layer. Therefore, it is beneficial to boost the magnitude of spin torque to enable switching at lower current and voltage levels. In contrast, for magnetic sensing, spin-transfer torques introduce noise and instability[7,8]. In this letter, we adapt a proposal due to Berger[9] to demonstrate that adding a copper spacer and a third magnetic fixed layer above the MTJ structure can, depending on the relative orientation of the two fixed layers, either increase by ~70% or nearly cancel the net spin-torque acting on the free magnetic layer. We analyze results by considering the effects of the spin-torque and of the substantial Joule heating that occurs in tunnel junction devices at switching current levels. This investigation provides fundamental insight into the nature of spin torques in MTJs and spin valves by allowing a direct comparison, in the same device, of the magnitude of the spin-torque exerted by tunnel currents and by electrons that transport spin through the copper spacer layer.

The multilayers for our samples were prepared on a thermally oxidized Si wafer in a DC magnetron sputtering chamber with a base pressure of $3\times10^{-8}$ torr. The layers consisted of (in nm): Py 4/Cu 20/Ta 5/Cu 20/Ta 20/CoFe 8/AlO$_x$ ~0.7-0.8/Py 4/Cu 6/CoFe 5/Cu 5/Pt 30, where Py is Ni$_{81}$Fe$_{19}$ and CoFe is Co$_{86}$Fe$_{14}$. The AlO$_x$ layer is formed from the room temperature thermal oxidation of a 0.65 nm layer of Al with an oxygen dose of 27 Torr-s. The multilayers were patterned by a combination of e-beam lithography and ion milling into elliptically shaped (~ 35 × 90 nm) CPP nanopillars with Cu top contacts. The 4 nm of Py above the AlO$_x$ serves as the magnetic free layer and the two CoFe layers are both magnetic fixed layers in our studies. Since the tunnel barrier limits the conductance, the resistance state reflects whether the bottom fixed layer (8 nm CoFe) and the free layer are aligned parallel (P, low resistance) or antiparallel (AP, high resistance). The role of the top fixed layer (5 nm CoFe) is to provide an additional source of spin-polarized electrons to the free layer. When the two fixed layers are anti-aligned with respect to one another (Fig. 1(a)), the spin-currents from both fixed layers exert torque in the same direction. When the fixed layers are aligned, (Fig. 1(b)) the spin-polarized electrons from the two fixed layers exert torque in opposite directions, resulting in a much smaller net torque.

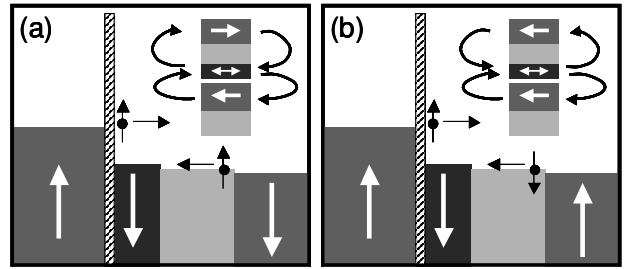

FIG. 1. Electron energy diagram for anti-aligned fixed layers (a) and aligned fixed layers (b) for a net electron current flowing from left to right. Arrows on the electrons indicate the direction of the spin current due to each fixed layer. Inset (a) & (b): Diagrams of how the fixed layer magnetic configuration affects the net dipole field of the free layer.

After setting the alignment of the two fixed magnetic layers, we characterize the properties of the free layer nanomagnet by simultaneously measuring differential resistance ($dV/dI$) and DC resistance ($R_{DC}$) while sweeping magnetic field ($H$) at a fixed current bias ($I$), as well as sweeping $I$ at fixed $H$. Measurements were made at $T$=77 K since the small free layer is not thermally stable at room temperature. Figure 2(a) and (b) show representative sweeps of $H$ and $I$. This sample has a tunnel magnetoresistance (TMR) of 6.6% at 77 K, and 4.3% at room temperature. From these data, which are typical of a number of samples studied, we construct a free-layer phase diagram for the two magnetic alignments of the fixed layers, shown in Figure 2(c) and (d). These phase diagrams demonstrate that both spin-torque and Joule heating play important roles in the thermally activated switching of these samples. When fixed layers are aligned (Fig. 2(c)), the observation of a phase diagram that is nearly symmetric about $I=0$ suggests immediately that spin-torque effects are weak and heating is dominant for this configuration. The power dissipated in MTJs at similar current densities has previously been observed to cause substantial heating[10], in contrast to metallic spin-valve nanopillars where heating effects are small[11].

To model the effects of spin torque and heating on the switching behavior we begin with the Sharrock formula[12] for the coercive field $H_{sw}$ in a thermal activation picture. We then modify the energy-activation barrier by the factor $(1-I/I_{c,o})$ to account for the effects of spin-transfer torque, predicted theoretically[13,14] and observed experimentally[11,15,16] in spin valve nanopillars well below the $T=0$ critical current $I_{c,o}$. We obtain,

$$\left|H_{sw} - H_{dip}\right| = H_{c,o}(T)\left(1 - \sqrt{\frac{k_B T}{E_a(T)}\left(1-\frac{I}{I_{c,o}}\right)^{-1} \ln\left(\frac{\tau_m}{\tau_o \ln 2}\right)}\right), \quad (1)$$

where $H_{dip}$ is the external field needed to cancel the average dipole field acting on the free layer, $H_{c,o}$ is the magnitude of the $T=0$ coercive field, $\tau_m$ is the measurement time per step (in our case, 1 s), $\tau_o=10^{-9}$ s is the attempt time[11], and $E_a$ is the activation barrier. $H_{dip}$ = 742 Oe when the two fixed layers are parallel and 272 Oe when they are antiparallel, as determined from the midpoint of the free layer's minor loop (e.g., Fig. 2(a)). From fits to $H_{sw}$ as a function of $T$ near 77 K, we determine $E_a = 0.49$ eV and $H_{c,o} = 130$ Oe when the fixed layers are aligned, while $H_{c,o} = 110$ Oe when they are antiparallel. We attribute this difference to variations in reversal modes for the free layer under the influence of the non-uniform dipole fields from the fixed layers[17]. In our heating analysis, we will assume that $H_{c,o}$ and $E_a$ have $T$ dependences approximately proportional to the $M(T)$ and $[M(T)]^2$, respectively, where $M(T)$ is the magnetization we measure with SQUID magnetometry for a superlattice of 4 nm Py layers with Cu.

To account for Joule heating, we model the device as a one-dimensional inhomogeneous wire with local heating at the tunnel junction and a thermal conductivity dominated by conduction electrons and obeying the Weidemann-Franz law. By integrating the heat flow equation, we estimate that the temperature at the free layer is:

$$T = \sqrt{T_{bath}^2 + \beta I^2}. \quad (2)$$

Here $\beta$ is a parameter that depends on material specific factors and device geometry. Using Eq. (1) and (2), we make independent fits to the $H_{sw}$ data for P→AP switching and for AP→P switching using $\beta$ and $I_{c,o}$ as fitting parameters.

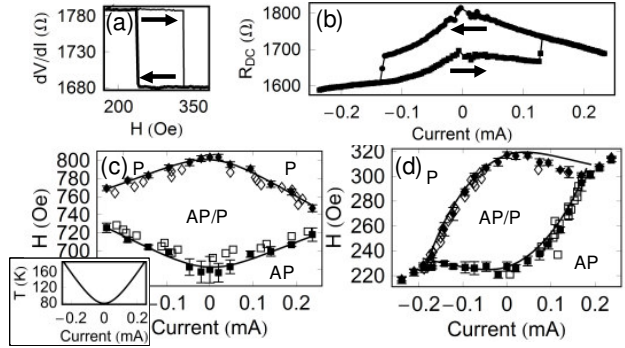

FIG. 2. (a) Sweep of $H$ for anti-aligned fixed layers measured without DC current bias and ~2 A of AC sense current. (b) Sweep of $I$ for anti-aligned fixed layers measured at a constant $H = 278$ Oe. Squares (■) are for increasing $I$ and circles (●) are for decreasing $I$. Phase diagrams for aligned (c) and anti-aligned (d) fixed layers, constructed by sweeping $H$ at fixed $I$ and by sweeping $I$ at fixed $H$ for each magnetic configuration. States are represented by alignment between the bottom fixed layer and the free layer; either parallel (P) or anti-parallel (AP). For $H$ sweeps, P to AP transitions are marked by filled squares (■) while AP to P is marked with filled diamonds (♦). For $I$ sweeps, P to AP transitions are marked by open squares (□), while AP to P is marked with open diamonds (◊). Lines represent fits to Eqs. (1) and (2). All measurements shown were made at 77 K. Inset: Calculated $T$ of the free layer as function of $I$.

Beginning with the aligned fixed layer phase diagram (Fig. 2(c)), we were able to obtain good fits to the P→AP and AP→P data with the independently-determined heating parameters, $\beta_{P\rightarrow AP} = (4.4\pm0.3) \times 10^5$ K$^2$/mA$^2$ (in terms of power density,[18] 0.65±0.14 K$^2$ μm$^2$/μW) and $\beta_{AP\rightarrow P} = (4.7\pm0.3) \times 10^5$ K$^2$/mA$^2$ (0.71±0.14 K$^2$ μm$^2$/μW). The similarity of these independent parameters demonstrates that the model accounts consistently for heating. The inset in Fig. 2 (c) shows $T(I)$ as calculated from Eq. (2) and $\beta = 4.7 \times 10^5$ K$^2$/mA$^2$. From the fits we also extract values of $I_{c,o\ P\rightarrow AP} = -2.8\pm1.0$ mA and $I_{c,o\ AP\rightarrow P} = 1.4\pm0.2$ mA for the aligned-fixed-layers case, corresponding to $T=0$ critical current densities normalized by free layer thickness of $(J_{c,o}/t)_{P\rightarrow AP} = (-2.8\pm1.0) \times 10^7$ A/(cm$^2$ nm) and $(J_{c,o}/t)_{AP\rightarrow P} = (1.4\pm0.2) \times 10^7$ A/(cm$^2$ nm). The same values measured in a metallic spin valve with a Py free layer[11] at $T=4.2$ K are $(J_c/t)_{P\rightarrow AP} \approx 6.6 \times 10^6$ A/(cm$^2$ nm) and $(J_c/t)_{AP\rightarrow P} \approx -3.3 \times 10^6$ A/(cm$^2$ nm). We conclude that the cancellation of the torques from the two fixed layers reduces the net torque by a factor of ~4. Since the sign of the residual spin-torque is reversed from that seen in normal MTJs[6], where $+I$ favors AP in our convention, the spin-torque exerted by electrons transported though the copper spacer layer is slightly stronger than the torque from the tunneling electrons in these devices.

With $\beta$ obtained from the aligned fixed layer data, we fit to the $H_{sw}$ data taken with anti-aligned fixed layers (Fig. 2(d)), to determine $I_{c,o\ P\rightarrow AP}$ and $I_{c,o\ AP\rightarrow P}$, employing the same device parameters as in the aligned case, except that here $H_{c,o} = 110$ Oe, instead of 130 Oe, as explained above. We obtain[19] $I_{c,o\ P\rightarrow AP} = 0.29\pm0.01$ mA and $I_{c,o\ AP\rightarrow P} = -0.28\pm0.01$ mA, and therefore $(J_{c,o}/t)_{P\rightarrow AP}$

= (2.9±0.6) × 10$^6$ A/(cm$^2$ nm) and $(J_{c,o}/t)_{AP \rightarrow P}$ = (-2.8±0.6) × 10$^6$ A/(cm$^2$ nm). We have previously noted that $(J_c/t)_{P \rightarrow AP} \approx 6.6 \times 10^6$ A/(cm$^2$ nm) and $(J_c/t)_{AP \rightarrow P} \approx -3.3 \times 10^6$ A/(cm$^2$ nm) and therefore $(J_{c,o}/t)_{P \rightarrow AP}$ decreases by 60%, and $(J_{c,o}/t)_{AP \rightarrow P}$ decreases by 15% relative to the single fixed layer case. This corresponds to a 20-130% increase in the strength of the torque.

Slonczewski has predicted the spin-torque per unit area $N_{sv}$, exerted by the current in a spin valve[20] and also $N_{tunnel}$ from the tunnel current in a MTJ[21]. The spin valve result is $N_{sv} = (\hbar/e)g_{sv}(P,\theta)J\sin\theta$, where $P$ is the polarization induced in a current passing through either of the ferromagnetic layers, $J$ is the current density, $\theta$ is the angle between the magnetizations of the electrodes, and $g_{sv}(P,\theta) = [-4 + (1+P)^3(3+\cos\theta)/(4P^{3/2})]^{-1}$. For the tunnel case, $N_{tunnel} = (\hbar/e)g_{tunnel}(P_{tunnel},\theta)J\sin\theta$ where $P_{tunnel}$ is the polarization of the tunnel current, assumed the same for currents from either electrode, and $g_{tunnel} = (1/2)P_{tunnel}/(1+P_{tunnel}^2 \cos\theta)$. Due to the short spin relaxation length in Py, we neglect possible spin accumulation effects,[9,22,23] and simply assume $I_{c,o} \propto 1/N_{net}$, where $N_{net} = N_{sv} \pm N_{tunnel}$, with the sign depending on the fixed layer alignment. Then, the ratio of the critical currents for the two cases would be:

$$\rho(\theta) = \frac{I_{c,o}^{Aligned}(\theta)}{I_{c,o}^{Anti-aligned}(\theta)} = \frac{g_{sv}(\pi-\theta) + g_{tunnel}(\theta)}{|g_{sv}(\theta) - g_{tunnel}(\theta)|} \quad (3)$$

where for the anti-aligned case the argument of $g_{sv}$ is $(\pi-\theta)$ since $\theta$ is measured with respect to the bottom fixed layer. Using the Julliere model[24] and the TMR of our MTJs we can estimate the polarization of the tunnel current, $P_{tunnel}$~18%. If we assume that the polarization of a current passing through the CoFe layer is $P$~35%, similar to values of polarization measured in Co[25], from Eq. (3) we calculate $\rho(\theta=0)$ = 14.9 and $\rho(\theta=\pi)$ = 0.5. Our measured values are $\rho(\theta=0)$ = 9.7±3.4 and $\rho(\theta=\pi)$ = 5.0±0.7, which indicates a qualitatively different behavior, especially for $\theta=\pi$. The difference could be due either to substantial spin accumulation[9,22,23] effects in this compound structure, or to the strength and angular dependence of the torque differing from that predicted from the TMR by an ideal tunneling model[21,26].

In summary, by addition of a second fixed magnetic layer, the spin torque exerted by a current on the free layer of an MTJ structure can, depending on the magnetic alignment of the fixed layers, be substantially enhanced or reduced. Both can be beneficial depending on the MTJ application. We have exploited the ability to nearly cancel the spin-torque in our devices to quantify self-heating effects in MTJs at high bias where spin-torque switching is dominant. For practical MRAM, devices would need a thicker (~8nm) free layer to provide thermal stability at room temperature ($E_a > 1\ eV$). While this would push the $T=0$ switching current to ~0.6 mA, this is within the range that a scaled CMOS transistor can source. We note here that for high-speed MRAM, pulsed currents of ns duration will be employed for writing in the torque-driven regime[16,27], and thus it is the $T = 0$ switching current that is the relevant value. However, the substantial self-heating may somewhat reduce the required pulse amplitude due to magnetization reduction. Finally, the spin torque due to tunneling in our devices is larger than predicted from the low TMR (6.6%) and has approximately the same magnitude as the torque from spin-polarized electrons passing through the copper spacer layer. If the prediction that $I_{c,o}$ in MTJs decreases with increasing TMR[21], and if MTJs with much higher TMR and sufficiently low $RA$ can be engineered, then the possibility is open for much lower critical current densities. This is encouraging for the future development of spin-transfer switched MRAM.

This research was supported by the Army Research Office, by the NSF/NSEC program through the Cornell Center for Nanoscale Systems, and by DARPA through Motorola, and benefited from use of the NSF-supported Cornell Nanoscale Science and Technology Facility/NNIN and the facilities of the Cornell Center for Materials Research.

[1] S. Tehrani *et al., Proc. Inst. Electr. Elect.* **91**, 703 (2003).
[2] J. DeBrosse, *et al., IEEE J. Solid-St. Circ.* **39**, 678 (2004).
[3] J. A. Katine, F. J. Albert, E.B. Myers, D. C. Ralph, and R. A. Buhrman, *Phys. Rev. Lett.* **84**, 3149 (2000).
[4] F. J. Albert, J. A. Katine, D. C. Ralph, and R. A. Buhrman, *Appl. Phys. Lett.* **77**, 3809 (2000).
[5] Y. Huai, F. Albert, P. Nguyen, M. Pakala, and T. Valet. *Appl. Phys. Lett.* **84**, 3118 (2004).
[6] G. D. Fuchs *et al., Appl. Phys Lett.* **85**, 1205 (2004).
[7] A. V. Nazarov, H. S. Cho, J. Nowak, S. Stokes, and N. Tabat, *Appl. Phys. Lett.* **81**, 4559 (2002).
[8] V. Synogatch, N. Smith, and J. R. Childress, *J. Appl. Phys.* **93**, 8570 (2003).
[9] L. Berger, J. *Appl. Phys.* **93**, 7693 (2003).
[10] R. C. Sousa *et al., J. Appl. Phys.* **95**, 6783 (2004).
[11] I. N. Krivorotov *et al., Phys. Rev. Lett.* **93**, 166603 (2004).
[12] M. P. Sharrock, *IEEE Trans. Magn.* **26**, 193 (1990).
[13] Z. Li and S. Zhang, *Phys. Rev. B* **69**, 134416 (2004).
[14] N. Smith, cond-mat/0406486 (2004).
[15] E. B. Myers *et al.*, *Phys. Rev. Lett.* **89**, 196801 (2002).
[16] R. H. Koch, J. A. Katine, and J. Z. Sun. *Phys. Rev. Lett.* **92**, 088302 (2004).
[17] R. H. Victora, *Phys. Rev. Lett.* **63**, 457 (1989).
[18] Scaled using device area, $A$=(2.5±0.5) x 10$^{-3}$ µm$^2$ and P resistance, $R$=1680 Ω.
[19] If we use a 2/3 power law rather than 1/2 (see, for example, ref. 17) in Eq. (1), we obtain values for $I_{c,o}$ that are 13% larger.
[20] J. C. Slonczewski, *J. Magn. Magn. Mater.* **159**, L1 (1996).
[21] J. C. Slonczewski, unpublished (2004).
[22] J. Xiao, A. Zangwill, and M. D. Stiles, cond-mat/0407569 (2004).
[23] J. Manschot, A. Brataas, and G. E. W. Bauer, *Appl. Phys. Lett.* **85,** 33250 (2004).
[24] M. Julliere, *Phys. Lett.* **54A**, 225 (1975).
[25] S. K. Upadhyay, R. N. Louie, and R. A. Buhrman, *Appl Phys. Lett.* **74**, 3881 (1999).
[26] J. C. Slonczewski, *Phys. Rev. B.* **71**, 024411 (2005).
[27] J. Z. Sun, *Phys. Rev. B.* **62**, 570 (2000).